\PassOptionsToPackage{safe}{tipa}
\documentclass[cameraready]{Interspeech}

\title{\method{}: Harmonic-Aware Residual Partitioning for Neural Audio Codecs}

\author[affiliation={1}]{Qiaoyu}{Yang}
\author[affiliation={2,3}]{Lixing}{He}
\author[affiliation={1}]{Binyue}{Deng}
\author[affiliation={3}]{Weifeng}{Zhao}

\address{
    $^1$ Georgia Institute of Technology, Atlanta, United States \\
    $^2$ The Chinese University of Hong Kong, Hong Kong, China \\
    $^3$ Tencent Music Entertainment, Shenzhen, China
}

\email{qyang312@gatech.edu}


\usepackage{comment}

\DeclareMathOperator*{\argmin}{arg\,min}

\newcommand{\method}{HARP}

\usepackage{amsmath}
\usepackage{amssymb}
\usepackage{booktabs}
\usepackage{multirow}
\usepackage{algorithm}
\usepackage{algorithmic}

\begin{document}

\maketitle

\begin{abstract}
Neural audio codecs with residual vector quantization (RVQ) normally treat all frequencies uniformly, so their codebooks become spectrally entangled. Truncating stages then removes an unpredictable mix of frequencies. Parallel band decomposition addresses this by splitting audio into independent bands, but fragments the latent space and loses cross-frequency coherence. We introduce HARP (Harmonic-Aware Residual Partitioning), a training strategy that partitions RVQ stages into frequency-ordered groups where each group refines its target band while the decoder retains access to all lower frequencies. Overtones are reconstructed in the context of their fundamentals, preserving coherence that parallel methods lose. HARP requires no architectural changes; it only modifies the training loss, leaving inference identical to standard RVQ. On speech, music, and general audio, HARP outperforms both standard RVQ and parallel decomposition. MUSHRA listening tests also show perceptual improvements.
\end{abstract}

\section{Introduction}

\begin{figure*}[t]
  \centering
  \includegraphics[width=\linewidth]{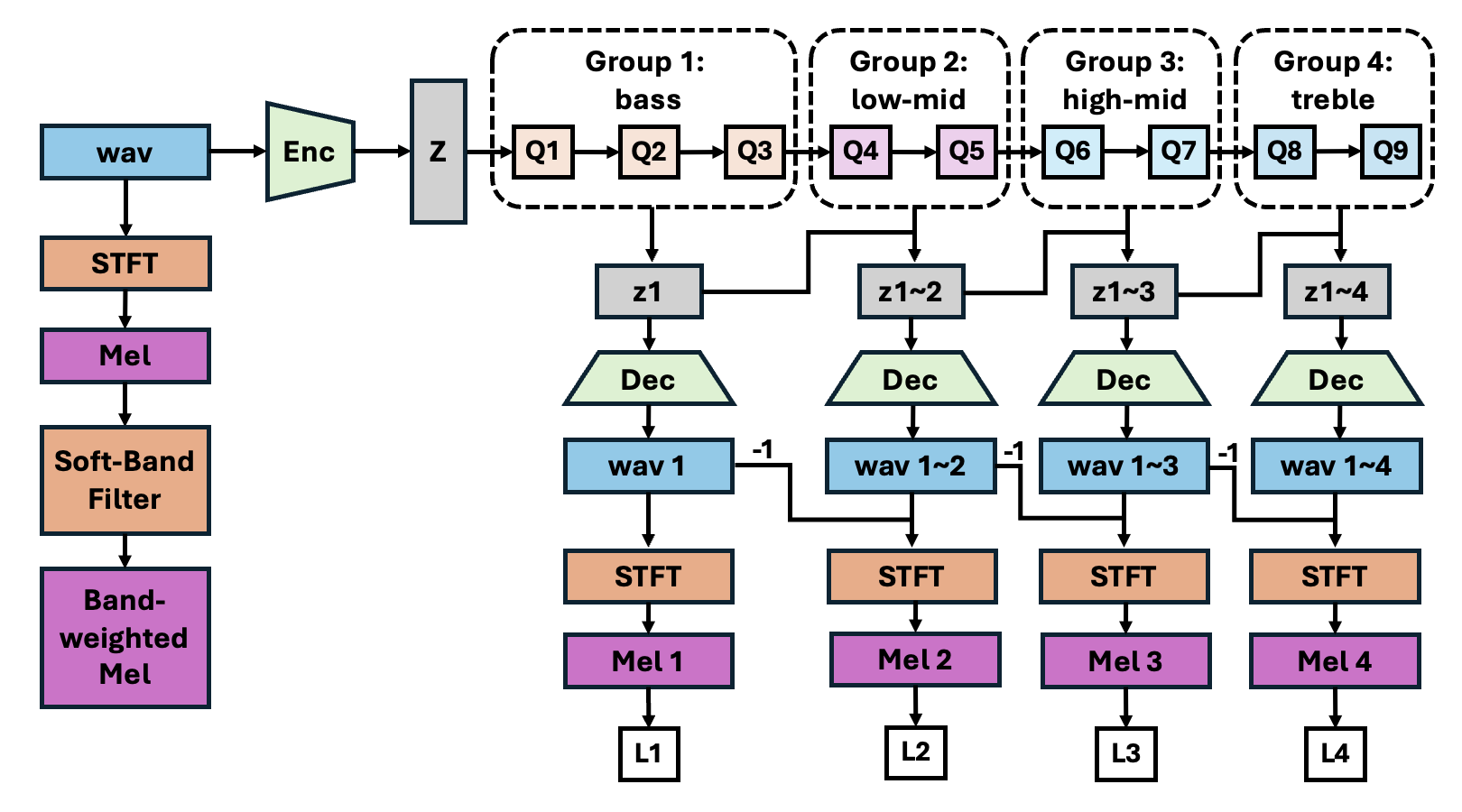}
  \caption{Overview of the \method{} pipeline. Nine RVQ stages are partitioned into four groups targeting progressively higher frequency bands. At each group boundary, the cumulative latent ($\hat{z}_{\leq k}$) is decoded and its mel-spectrogram contribution is supervised against a band-weighted reference ($-1$ denotes subtraction of the previous group's output). At inference, only a single encoder--decoder forward pass is needed, identical to standard RVQ.}
  \label{fig:overview}
\end{figure*}

Neural audio codecs~\cite{zeghidour2021soundstream, defossez2022highfidelity, kumar2023highfidelity} compress speech, music, or environmental sound directly from data, matching or exceeding classical codecs at low bitrates. Beyond compression, they produce discrete tokens that bridge continuous audio and language models, enabling speech synthesis~\cite{wang2023neural}, music generation~\cite{copet2024simple,borsos2023audiolm}, and audio understanding.

Most modern neural audio codecs rely on residual vector quantization (RVQ): each stage quantizes the residual left by earlier stages, progressively refining a single latent space. This provides flexible bitrate control (more stages, higher fidelity) and a token stream well suited to autoregressive modeling, all within a single encoder--decoder pair.

RVQ, however, is frequency-agnostic. Each stage captures whatever content reduces residual error most, without regard for spectral structure. This becomes apparent when truncating a pretrained DAC model to fewer stages for low-bitrate operation: the resulting artifacts vary unpredictably, sometimes removing bass entirely, sometimes treble, with no consistency across inputs. Computing spectral centroids for each of the nine codebook stages reveals that they all cluster in a narrow range with high variance. Stage~1 covers roughly the same frequency mix as stage~9. We refer to this as \emph{spectral entanglement}. It renders the representation difficult to interpret and, more practically, makes bitrate scaling unpredictable.

This is consequential because audio is far from spectrally flat. Bass (20--250~Hz) dominates signal energy; midrange (250~Hz--4~kHz) carries speech intelligibility and melody; treble (4--20~kHz) adds brightness and texture. A codec should prioritize perceptually foundational content in early stages. Furthermore, frequency bands are not independent: a violin at 440~Hz produces harmonics at 880, 1320, and 1760~Hz, all phase-locked. Disrupting that coherence yields hollow timbre. Speech exhibits the same property: vowel identity depends on the amplitude relationship between a fundamental and its overtones.

Parallel band decomposition~\cite{wang2025bscodec} addresses spectral structure by splitting audio into frequency bands via filterbanks, encoding each with a separate network, and quantizing independently. This achieves band-level specialization but fragments the representation: multiple encoders, multiple token streams, and isolated decoders. When reconstructing midrange, the decoder has no access to the bass content. Harmonics that span bands must be recombined implicitly, with no guarantee of phase or amplitude coherence.

We propose a third path. \method{} (\textbf{H}armonic-\textbf{A}ware \textbf{R}esidual \textbf{P}artitioning) trains RVQ stages to form a frequency hierarchy \emph{within} the standard architecture: early stages capture bass, middle stages add midrange, and later stages refine treble. The term \emph{harmonic-aware} refers to the preservation of cross-band phase and amplitude relationships through cumulative decoding, rather than an assumption of musically harmonic input; the mechanism applies equally to polyphonic music, inharmonic sounds, and noise-like content, since energy concentration in low-to-mid bands is a property of natural audio in general. The core mechanism is straightforward: each stage group refines its target band while the decoder retains access to all lower-frequency content. When learning to produce midrange, the decoder receives the bass latent, so a 200~Hz fundamental is available to guide coherent reconstruction of its overtones. We term this \emph{cumulative decoding}.

Two additional training-time mechanisms complement cumulative decoding. First, we supervise each group's \emph{contribution}, the spectral improvement it adds over previous groups, rather than the cumulative output, keeping each group focused on its designated frequency region (\emph{subband contribution supervision}). Second, we apply learnable Gaussian weights over mel bins to softly steer each group toward its target band, avoiding the sharp cutoff boundaries that empirically produce edge artifacts (\emph{soft band weighting}).

At inference, \method{} is identical to standard RVQ: no additional parameters or forward passes, and the same unified token stream. The hierarchical structure resides in the learned codebooks, not the architecture. Bitrate scaling becomes predictable. Dropping later stages removes treble before bass, so quality degrades gracefully rather than erratically.

\textbf{Contributions.} (1)~We characterize spectral entanglement in standard RVQ and harmonic incoherence in parallel band decomposition as complementary failure modes. (2)~We propose \method{}, a training-only strategy that imposes frequency hierarchy on RVQ through cumulative decoding and soft subband supervision, requiring no architectural changes. (3)~We demonstrate that \method{} outperforms both standard RVQ and parallel decomposition across speech, music, and general audio, with gains confirmed by objective metrics and MUSHRA listening tests at two bitrate conditions. Pre-trained models, training scripts, and audio samples are publicly available.\footnote{\url{https://github.com/QiaoyuYang/harp-codec}}

\section{Related Work}

\subsection{Neural Audio Codecs}

End-to-end neural audio compression pairs a learned encoder--decoder with a discrete bottleneck,
trained jointly with reconstruction and adversarial losses. SoundStream~\cite{zeghidour2021soundstream}
established the dominant recipe, a convolutional encoder--decoder with residual vector quantization
(RVQ) and multi-scale discriminators. Encodec~\cite{defossez2022highfidelity} follows Soundstream
with improved multi-objective loss balancing and DAC~\cite{kumar2023highfidelity} with periodic
Snake activations~\cite{ziyin2020snake} and stronger discriminators at 44.1~kHz.

Several lines of work extend this foundation along different axes. On the \emph{quantization} side,
HiFi-Codec~\cite{yang2023hifi} partitions the latent into groups with separate RVQ chains, and
WavTokenizer~\cite{ji2024wavtokenizer} replaces multi-stage RVQ with a single large codebook
aligned with semantic information. On the \emph{token design} side, SNAC~\cite{siuzdak2024snac}
operates quantizers at different frame rates to capture multi-scale structure,
FlexiCodec~\cite{li2025flexicodec} dynamically allocates frame rates based on content complexity,
and SpeechTokenizer~\cite{zhang2023speechtokenizer} distills semantic information from
self-supervised models into the first RVQ codebook, separating semantic content from acoustic
detail across stages. On the \emph{backbone architecture} side, FlowDec~\cite{welker2025flowdec}
replaces adversarial training with a conditional flow matching postfilter, matching GAN-based
quality on 48~kHz general audio with improved harmonic reconstruction, and
Parker et al.~\cite{parker2025scaling} scale a transformer-based codec to 1B parameters with an
FSQ bottleneck, achieving high-quality speech at 400--700~bps.

Across all of these designs, RVQ stages or their equivalents remain spectrally unstructured. No
stage is explicitly assigned to any frequency region.

\subsection{Quantization Strategies}

VQ-VAE~\cite{oord2017neural} popularized discrete latent bottlenecks for generative modeling.
RVQ~\cite{zeghidour2021soundstream} decomposes quantization across a cascade of codebooks, each
encoding the residual from the previous stage, yielding a coarse-to-fine representation and
flexible bitrate control. Product quantization~\cite{jegou2010product} partitions the latent
vector into subvectors quantized independently, and group-residual VQ~\cite{yang2023hifi} extends
this idea by assigning a separate RVQ chain to each group. SRCodec~\cite{zheng2024srcodec} further
refines the split: it divides the latent into two equal-dimension parts via an attention-driven
dual module, then quantizes the low-dimensional part and the residual between the two, improving
coding efficiency at low bitrates.

A recent line of work seeks to simplify or replace the VQ codebook entirely. Finite scalar
quantization (FSQ)~\cite{mentzer2023finite} quantizes each latent dimension independently to a
small set of fixed values, forming an implicit codebook as the Cartesian product of per-dimension
levels; this eliminates codebook collapse and the associated machinery of commitment losses,
reseeding, and entropy penalties. Binary spherical quantization (BSQ)~\cite{zhao2025bsq} projects
embeddings onto a hypersphere and applies binary quantization, yielding parameter-free codebooks
that scale to arbitrary token dimensions with high compression ratios. SimVQ~\cite{zhu2025simvq}
reparameterizes codebook vectors through a learnable linear transformation over a latent basis,
ensuring all codes receive gradient updates and maintaining near-complete utilization even at
codebook sizes exceeding 200k entries.

These quantization strategies address codebook capacity, utilization, and scalability, but none
impose spectral ordering on RVQ stages. \method{} is orthogonal to the choice of quantizer variant
and can in principle be applied atop any RVQ-based codec.

\subsection{Frequency-Aware Neural Audio Compression}

Several neural codecs depart from full-band processing by introducing explicit frequency structure
into the encoding pipeline.

\textbf{Parallel band decomposition.} BSCodec~\cite{wang2025bscodec} partitions the input
spectrogram into frequency bands via fixed binary masks, then passes each band through a dedicated
encoder--decoder and quantizer, yielding parallel token streams. Reconstruction is performed
independently per band, so harmonics that span band boundaries must be recombined without direct
cross-band interaction, and the resulting multi-stream bitrate complicates autoregressive language
modeling. SPCODEC~\cite{wen2025spcodec} adopts a lighter design: the latent embedding is split
along the channel dimension into low- and high-frequency components, with the low-frequency
component supervised by a reconstruction loss and the high-frequency component decorrelated from
it via an attention-based prediction module to reduce bitrate. A single shared decoder reconstructs
from both components, preserving cross-band access, though two codebook streams are still produced.

\textbf{Spectral-domain codecs.} TF-Codec~\cite{jiang2023latent} operates on a complex STFT
spectrogram and introduces a learnable scalar power-law compression on the amplitude, which shifts
the encoder's effective attention between dominant and fine-grained spectral components as a
function of bitrate, without partitioning the spectrum into separate processing paths.
APCodec~\cite{ai2024apcodec} encodes amplitude and phase spectra through parallel branches for
improved high-frequency reconstruction. Both approaches embed frequency sensitivity into the input
representation rather than into the quantization structure.

\textbf{Frequency disentanglement.} \cite{ginies2025soft} trains a
two-branch codec at 16~kHz and 32~kHz, where the second branch encodes the residual signal left by
the first. Frequency hierarchy is imposed through the branching architecture rather than through
modifications to the training objective of a shared model.

\method{} differs from all of these in that frequency specialization emerges within a single
encoder, decoder, and token stream, with no architectural changes; only the training loss is
modified. Cumulative decoding additionally ensures that each successive RVQ stage has access to
lower-frequency reconstructions, preserving harmonic structure that band-partitioned designs
sacrifice by construction.

\subsection{Perceptual Coding and Psychoacoustic Bit Allocation}

Classical audio codecs distribute coding capacity according to psychoacoustic models of human
auditory perception. MP3~\cite{painter2000perceptual} and AAC exploit critical-band analysis,
simultaneous masking, and temporal masking to concentrate precision in perceptually salient
regions and withhold bits from masked ones. Opus~\cite{valin2012definition} pairs a
linear-predictive coder (SILK) with an MDCT-based transform coder (CELT), selecting between them
based on signal characteristics. Across these designs, a consistent empirical finding motivates
the architecture: perceptual salience is strongly non-uniform across frequency, with low and mid
frequencies dominating speech intelligibility and musical pitch, and high frequencies contributing
timbral texture that tolerates coarser quantization.

End-to-end neural codecs have largely replaced explicit psychoacoustic models with learned
perceptual surrogates---multi-resolution STFT losses, mel-spectrogram distances, and adversarial
objectives. MUFFIN~\cite{ng2025muffin} departs from this trend by introducing Multi-Band Spectral
RVQ (MBS-RVQ), which applies an FFT to the encoder's latent representation to isolate frequency
bands, then allocates quantization capacity across bands according to psychoacoustic salience.
Separate codebooks are used for content and speaker information. Unlike the band-split
architectures, frequency partitioning in MUFFIN occurs entirely within the quantization bottleneck
and operates on latent features rather than the input spectrogram, leaving the encoder and decoder
architecturally unmodified. Without such explicit allocation, the distribution of information
across standard RVQ stages is not constrained to reflect perceptual importance.

\method{} recovers frequency-prioritized allocation through supervision on the mel scale during
training, without modifying codec architecture or hand-specifying psychoacoustic targets.
\section{Background}

\subsection{Neural Audio Codec Architecture}

A neural audio codec comprises an encoder $E$, a quantizer $\mathcal{Q}$, and a decoder $G$.
Given a waveform $x \in \mathbb{R}^T$, the encoder produces a continuous latent representation
$z = E(x) \in \mathbb{R}^{D \times T'}$, where $T' = T/S$ is determined by the cumulative stride
$S$ of the convolutional backbone (typically $S \in \{320, 512\}$ at 24--44.1~kHz). The decoder
reconstructs a waveform from the quantized latent: $\hat{x} = G(\hat{z})$. Both encoder and
decoder are built from residual convolutional blocks, with strided convolutions handling temporal
downsampling and upsampling respectively.

A standard choice of nonlinearity for audio generation is the Snake activation~\cite{ziyin2020snake},
\begin{equation}
    \mathrm{Snake}_\alpha(x) = x + \frac{\sin^2(\alpha x)}{\alpha},
\end{equation}
which is periodic and monotonic, with a learnable frequency parameter $\alpha$ that controls the
rate of the periodic modulation. Originally proposed for learning periodic functions in general
regression settings, Snake was adopted for neural audio codecs by DAC~\cite{kumar2023highfidelity},
where the harmonic and quasi-periodic structure of audio makes a periodic inductive bias
particularly well-suited.

\subsection{Residual Vector Quantization}

Vector quantization (VQ), as used in VQ-VAE~\cite{oord2017neural}, discretizes a continuous
vector by mapping it to the nearest entry in a learned codebook
$\mathcal{C} = \{c_1, \ldots, c_N\} \subset \mathbb{R}^d$:
\begin{equation}
    \mathcal{Q}(z) = \argmin_{c \in \mathcal{C}} \lVert z - c \rVert_2.
\end{equation}
A single codebook of practical size cannot represent the full range of audio latents at acceptable
fidelity. Residual vector quantization (RVQ)~\cite{zeghidour2021soundstream} addresses this by
applying $L$ VQ stages sequentially, each operating on the residual left by the previous stage:
\begin{align}
    r^{(1)} &= z, \\
    q^{(\ell)} &= \mathcal{Q}^{(\ell)}(r^{(\ell)}), \\
    r^{(\ell+1)} &= r^{(\ell)} - q^{(\ell)},
\end{align}
with the final quantized representation $\hat{z} = \sum_{\ell=1}^{L} q^{(\ell)}$. At frame rate
$f$, using $L$ stages with $N$-entry codebooks yields a bitrate of $L \cdot \log_2(N) \cdot f$
bps. The sequential residual structure means that early stages capture coarse signal structure
while later stages refine progressively finer detail---a property that \method{} exploits through
targeted supervision.

\subsection{Mel-Spectrograms}
\label{sec:mel}

The mel-spectrogram maps a power spectrum to a compact perceptual representation via a mel-scaled
filterbank,
\begin{equation}
    \mathcal{M}(x) = \log\!\left(\mathbf{W}_{\text{mel}} \cdot |\mathrm{STFT}(x)|^2 +
    \epsilon\right),
\end{equation}
where $\mathbf{W}_{\text{mel}} \in \mathbb{R}^{M \times F}$ projects $F$ linear frequency bins
onto $M$ mel-scale bins. The mel scale allocates more bins to lower frequencies, approximating the
non-uniform frequency resolution of the cochlea and broadly reflecting the perceptual importance
of different spectral regions. We use $M = 80$ mel bins spanning 0--22~kHz. This perceptual
frequency weighting is central to the supervision strategy of \method{}, where losses are applied
at mel scale to encourage RVQ stages to specialize by perceptual frequency band.

\subsection{Training Objectives}
\label{sec:background}

Codec training combines reconstruction, adversarial, and quantization losses. The
\textbf{reconstruction loss} penalizes spectral differences at multiple resolutions,
\begin{equation}
    \mathcal{L}_{\text{rec}} = \sum_{s} \left( \lVert \Delta_s \rVert_1 + \lVert \Delta_s
    \rVert_2 \right) + \lVert \mathcal{M}(x) - \mathcal{M}(\hat{x}) \rVert_1,
\end{equation}
where $\Delta_s = \bigl||\mathrm{STFT}_s(x)| - |\mathrm{STFT}_s(\hat{x})|\bigr|$ is the
absolute magnitude difference at STFT scale $s$~\cite{yamamoto2020parallel}, and
$\mathcal{M}(\cdot)$ denotes the mel-spectrogram defined in Section~\ref{sec:mel}. The \textbf{adversarial loss} from multi-period and
multi-scale discriminators~\cite{kong2020hifi} encourages perceptual realism beyond what
reconstruction metrics alone capture.

Quantization introduces a non-differentiable argmin, through which gradients are passed via the
straight-through estimator~\cite{bengio2013estimating}. Two auxiliary losses align the encoder
and codebook. The \textbf{commitment loss} trains the encoder to stay close to its assigned
codeword,
\begin{equation}
    \mathcal{L}_{\text{commit}} = \sum_{\ell=1}^{L} \left\lVert r^{(\ell)} -
    \mathrm{sg}\left[q^{(\ell)}\right] \right\rVert_2^2,
\end{equation}
where $\mathrm{sg}[\cdot]$ is the stop-gradient operator, which prevents the loss from acting on
the codebook. The complementary \textbf{codebook loss} moves codebook entries toward the encoder
outputs,
\begin{equation}
    \mathcal{L}_{\text{cb}} = \sum_{\ell=1}^{L} \left\lVert \mathrm{sg}\left[r^{(\ell)}\right] -
    q^{(\ell)} \right\rVert_2^2,
\end{equation}
with the stop-gradient now on the encoder output so the loss acts only on the codebook. In
practice, the codebook loss has a closed-form solution equivalent to assigning each entry to the
mean of its assigned vectors; many codecs therefore replace it with exponential moving average
updates~\cite{zeghidour2021soundstream,defossez2022highfidelity,oord2017neural}, while others retain
explicit gradient-based codebook learning~\cite{kumar2023highfidelity}.

\section{Method}

\method{} modifies only the training objective. Architecture and inference are
identical to standard RVQ, incurring no additional cost at deployment.

\subsection{Design Motivation}

We seek to partition RVQ stages by frequency while preserving harmonic
coherence across bands. A natural first attempt---applying bandpass filters
directly to the supervision targets of each stage group---performs poorly:
rectangular frequency splits (zero weight outside each target band) introduce
discontinuities at band edges, and stages
in adjacent groups have no incentive to coordinate. We identify three
requirements for a viable solution.

\textbf{R1: Hierarchical specialization.} Early stages should capture bass
and later stages treble, consistent with both energy distribution (low
frequencies dominate) and perceptual priority (bass provides the harmonic
foundation).

\textbf{R2: Harmonic context.} Higher-frequency stages must have access to
lower-frequency content during training. A midrange partial at 800~Hz should
be reconstructed with knowledge that its 200~Hz fundamental is already
represented.

\textbf{R3: Soft boundaries.} Rectangular frequency splits---where the
supervision weight is set to zero outside the target band---produce
band-edge artifacts and prevent stages from capturing cross-band transients;
boundaries should emerge as learned preferences rather than rigid constraints.

Standard RVQ satisfies R2 and trivially satisfies R3 (no boundaries exist),
but does not exhibit R1. Parallel subband decomposition achieves R1 and can
satisfy R3, but violates R2 because each band's decoder operates on an
isolated representation. \method{} satisfies all three.

\subsection{Hierarchical Stage Groups}

We partition $L$ RVQ stages into $K$ ordered groups, each targeting a
frequency band from low to high. Let $S_k$ denote the stage indices assigned
to group $k$. Table~\ref{tab:groups} shows the default configuration for
$L{=}9$, $K{=}4$ with a 3-2-2-2 stage allocation. Three codebooks are
allocated to bass to reflect its energy dominance and perceptual importance.
These frequency ranges are approximate targets enforced through soft
supervision, not rigid binary constraints.

\begin{table}[h]
\centering
\caption{Default stage group configuration ($L{=}9$, $K{=}4$).}
\begin{tabular}{cccc}
\toprule
Group $k$ & Stages $S_k$ & Band & Frequency \\
\midrule
0 & 1--3 & Bass      & 0--1~kHz   \\
1 & 4--5 & Low-mid   & 1--4~kHz   \\
2 & 6--7 & High-mid  & 4--10~kHz  \\
3 & 8--9 & Treble    & 10--22~kHz \\
\bottomrule
\end{tabular}
\label{tab:groups}
\end{table}

\subsection{Cumulative Decoding}

To preserve harmonic context~(R2), each group's reconstruction is computed
from the \emph{cumulative} quantized latent rather than from that group's
contribution alone. Let $G$ denote the shared decoder and $q^{(\ell)}$ the
quantized output of the $\ell$-th RVQ stage. The cumulative latent through
group $k$ is:
\begin{equation}
\hat{z}_{\leq k} = \sum_{j=0}^{k} \sum_{\ell \in S_j} q^{(\ell)},
\end{equation}
with corresponding waveform $\hat{x}_{\leq k} = G(\hat{z}_{\leq k})$. We set
$\hat{z}_{\leq -1} = 0$ and $\hat{x}_{\leq -1} = 0$ as base cases. Because $G$ receives all
lower-frequency contributions when reconstructing group $k$, each higher band
is supervised in the presence of its harmonic context. This is the key distinction
from parallel decomposition, where each band's decoder operates on an isolated
latent.

\subsection{Subband Supervision}

To encourage spectral specialization~(R1), we supervise each group
individually using a frequency-weighted mel reconstruction loss applied to
that group's isolated waveform contribution. Let
\begin{equation}
\hat{z}_k = \sum_{\ell \in S_k} q^{(\ell)}
\end{equation}
denote group $k$'s latent increment, so that $\hat{z}_{\leq k} = \hat{z}_{\leq k-1} + \hat{z}_k$.

Applying the band loss directly to $\hat{x}_{\leq k} = G(\hat{z}_{\leq k})$
would create a gradient imbalance: since $\hat{z}_{\leq k}$ contains
$\hat{z}_{\leq k-1}$, gradients from $\mathcal{L}_{\text{band}}^{(k)}$ would
propagate into all prior groups through the encoder, causing group~0 to
accumulate encoder gradients from all $K$ band losses while group $K{-}1$
receives gradients only from its own. To prevent this, we apply stop-gradient
to $\hat{z}_{\leq k-1}$ inside the decoder call and subtract the stopped prior
output, isolating gradients entirely to group $k$'s latent increment
$\hat{z}_k$. Using the stop-gradient operator $\mathrm{sg}[\cdot]$ from
Section~\ref{sec:background}, define:
\begin{equation}
\hat{x}_k = G\bigl(\mathrm{sg}[\hat{z}_{\leq k-1}] + \hat{z}_k\bigr)
           - \mathrm{sg}[\hat{x}_{\leq k-1}],
\label{eq:xk}
\end{equation}
$\hat{x}_k$ is group $k$'s waveform increment: the change in decoded output
attributable solely to group $k$. The two stop-gradients work in concert ---
the first blocks gradient flow through $\hat{z}_{\leq k-1}$ inside the decoder,
and the second ensures the subtracted prior waveform $\hat{x}_{\leq k-1}$
contributes no gradient at all.
Since $G$ is nonlinear, $\hat{x}_k \neq G(\hat{z}_k)$; the prior latent is
retained as context for the decoder while being excluded from the gradient
path. Note that codebook entries are updated via the commitment loss
$\mathcal{L}_{\text{vq}}$ using straight-through estimation and are not
directly targeted by $\mathcal{L}_{\text{band}}^{(k)}$. All groups are jointly
updated by the standard reconstruction and adversarial losses computed on the
final output $\hat{x}_{\leq K-1}$.

\subsection{Soft Band Weighting}

To encourage spectral specialization without imposing rectangular cutoff boundaries~(R3),
we weight the band loss for group $k$ toward its target frequency range using
a normalized Gaussian over the $M$ mel bins defined in
Section~\ref{sec:mel}:
\begin{equation}
\tilde{w}_k(m) = \beta + (1 - \beta) \exp\!(-(m - \mu_k)^2 / 2\sigma_k^2),
\label{eq:weights_unnorm}
\end{equation}
where $\mu_k$ is a learnable center frequency (in mel bins), $\sigma_k$ a
learnable bandwidth, and $\beta \in (0,1)$ a fixed floor that retains weak
supervision across the full spectrum, preventing band-edge artifacts. Both
$\mu_k$ and $\sigma_k$ are initialized to divide the mel axis evenly, 
$\mu_k$ at the center of each target band and $\sigma_k = M/(3K)$. They are adapted during training, allowing each group's effective bandwidth to shift
toward the data distribution. We normalize to ensure consistent loss magnitude
across groups:
\begin{equation}
w_k(m) = \frac{\tilde{w}_k(m)}{\sum_{m'=1}^{M} \tilde{w}_k(m')}.
\label{eq:weights}
\end{equation}
The floor $\beta = 0.3$ is fixed. The band loss for group $k$ is then:
\begin{equation}
\mathcal{L}_{\text{band}}^{(k)} = \sum_{m=1}^{M} \sum_{t}
  w_k(m) \left| \mathcal{M}_{m,t}(\hat{x}_k)
  - \mathcal{M}_{m,t}(x) \right|,
\label{eq:band_loss}
\end{equation}
where $m$ indexes mel bins, $t$ indexes time frames, and $\hat{x}_k$
is defined in Eq.~\eqref{eq:xk}. Learning both $\mu_k$ and $\sigma_k$
allows each group's frequency preference to adapt in both center and width,
accommodating the unequal energy distribution across bands without requiring
manual tuning of band boundaries.

\subsection{Full Training Objective}

The complete loss combines standard neural codec terms with hierarchical band
supervision:
\begin{multline}
\mathcal{L} = \mathcal{L}_{\text{rec}} + \lambda_{\text{adv}}\mathcal{L}_{\text{adv}}
  + \lambda_{\text{fm}}\mathcal{L}_{\text{fm}} \\
  + \lambda_{\text{vq}}\mathcal{L}_{\text{vq}}
  + \lambda_{\text{band}} \sum_{k=0}^{K-1} \mathcal{L}_{\text{band}}^{(k)},
\end{multline}
where $\mathcal{L}_{\text{rec}}$ is the multi-scale mel reconstruction loss;
$\mathcal{L}_{\text{adv}}$ and $\mathcal{L}_{\text{fm}}$ are adversarial and
feature-matching losses from the discriminators; and $\mathcal{L}_{\text{vq}} =
\mathcal{L}_{\text{commit}} + \mathcal{L}_{\text{cb}}$ combines the commitment
and codebook losses from Section~\ref{sec:background} (or their EMA
equivalent)~\cite{zeghidour2021soundstream}. The reconstruction, adversarial,
and feature-matching losses operate on the full final output $\hat{x}_{\leq K-1}$.

\subsection{Group Dropout for Variable-Bitrate Operation}

To support variable-bitrate decoding, we apply group dropout during training.
With probability $p_{\text{drop}}$, the number of active groups $n$ is sampled
from $\text{Categorical}(\pi)$ with $\pi_n \propto n$ for $n \in \{1, \ldots, K\}$
(biased toward higher $n$); otherwise $n = K$. Active groups are always the lowest $n$ in the
hierarchy: bass stages are present in every training sample, while treble
stages are active only when all lower-frequency stages are also included. This
asymmetry reflects the role of each band---bass content is required at all
bitrates, whereas treble represents a higher-order refinement.

\subsection{Training and Inference}

Algorithm~\ref{alg:train} summarizes one training step. For each active group
$k$, the cumulative latent $\hat{z}$ is decoded to give $\hat{x}_{\text{curr}} = G(\hat{z})$.
The group increment $\hat{x}_k$ is then computed via a second decoder call
with stop-gradient applied to the prior cumulative latent $\hat{z}_{\leq k-1}$,
following Eq.~\eqref{eq:xk}. In the algorithm, $\hat{z}_{\leq k-1}$ is tracked
as $\hat{z}_{\text{prev}}$, so the second decoder call is
$G(\mathrm{sg}[\hat{z}_{\text{prev}}] + \hat{z}_k) - \mathrm{sg}[\hat{x}_{\text{prev}}]$,
where $\hat{z}_k = \hat{z} - \hat{z}_{\text{prev}}$ is group $k$'s latent increment.
The reconstruction, adversarial, feature-matching, and quantization losses are
computed on $\hat{x}_{\text{curr}}$ after the loop. That is, on the output
of the last active group, whether that is group $n{-}1$ under dropout or
$K{-}1$ at full rate. The band losses are computed at every group iteration
regardless of $n$.

\begin{algorithm}[t]
\caption{\method{} Training Step}
\label{alg:train}
\begin{algorithmic}[1]
\REQUIRE Waveform $x$, encoder $E$, decoder $G$,
         RVQ $\{\mathcal{Q}^{(\ell)}\}_{\ell=1}^L$
\REQUIRE Learnable centers $\{\mu_k\}_{k=0}^{K-1}$ and widths $\{\sigma_k\}_{k=0}^{K-1}$,
         floor $\beta$
\STATE $z \gets E(x)$
\STATE Sample $n \sim \text{Categorical}(\pi)$ w.p.\ $p_{\text{drop}}$,
       else $n \gets K$
\STATE $r \gets z$,\ $\hat{z} \gets 0$,\ $\hat{z}_{\text{prev}} \gets 0$,\
       $\hat{x}_{\text{prev}} \gets 0$,\ $\mathcal{L}_{\text{band}} \gets 0$
\FOR{$k = 0$ \textbf{to} $n-1$}
  \FOR{$\ell \in S_k$}
    \STATE $q^{(\ell)} \gets \mathcal{Q}^{(\ell)}(r)$;\
           $r \gets r - q^{(\ell)}$;\
           $\hat{z} \gets \hat{z} + q^{(\ell)}$
  \ENDFOR
  \STATE $\hat{z}_k \gets \hat{z} - \hat{z}_{\text{prev}}$
  \STATE $\hat{x}_{\text{curr}} \gets G(\hat{z})$
  \STATE $\hat{x}_k \gets G\bigl(\mathrm{sg}[\hat{z}_{\text{prev}}] + \hat{z}_k\bigr) - \mathrm{sg}[\hat{x}_{\text{prev}}]$
  \STATE Compute $w_k$ via Eqs.~\eqref{eq:weights_unnorm}--\eqref{eq:weights}
  \STATE $\mathcal{L}_{\text{band}} \mathrel{+}=
         \sum_{m,t} w_k(m)
         \left|\mathcal{M}_{m,t}(\hat{x}_k)
         - \mathcal{M}_{m,t}(x)\right|$
  \STATE $\hat{z}_{\text{prev}} \gets \hat{z}$;\
         $\hat{x}_{\text{prev}} \gets \hat{x}_{\text{curr}}$
\ENDFOR
\STATE Compute $\mathcal{L}_{\text{rec}},\, \mathcal{L}_{\text{adv}},\,
       \mathcal{L}_{\text{fm}},\, \mathcal{L}_{\text{vq}}$ on
       $\hat{x}_{\text{curr}}$
\RETURN $\mathcal{L}_{\text{rec}} + \lambda_{\text{adv}}\mathcal{L}_{\text{adv}}
        + \lambda_{\text{fm}}\mathcal{L}_{\text{fm}}
        + \lambda_{\text{vq}}\mathcal{L}_{\text{vq}}
        + \lambda_{\text{band}}\mathcal{L}_{\text{band}}$
\end{algorithmic}
\end{algorithm}

At inference, \method{} reduces to standard RVQ. Each stage output
$q^{(\ell)} = \mathcal{Q}^{(\ell)}(r^{(\ell)})$ is computed sequentially,
where $r^{(\ell)}$ is the residual entering stage $\ell$, and the waveform
is reconstructed as:
\begin{equation}
\hat{x} = G\bigl(\textstyle\sum_{\ell=1}^{L} q^{(\ell)}\bigr).
\end{equation}
The band parameters $\{\mu_k, \sigma_k\}$ are not used at inference;
hierarchical specialization is encoded entirely in the learned codebooks.
Variable-bitrate decoding is achieved by truncating to the first $k{+}1$
groups for any $k \in \{0, \ldots, K-1\}$:
\begin{equation}
\hat{x}_{\leq k} = G\bigl(\textstyle\sum_{j=0}^{k}\sum_{\ell \in S_j} q^{(\ell)}\bigr).
\label{eq:vbr}
\end{equation}

\textbf{Training cost.} The additional decoder forward passes for
subband contribution supervision (Eq.~\eqref{eq:xk}) increase training time
by approximately $2$--$3\times$ relative to standard DAC on the same hardware
(4$\times$ A100 GPUs in this work). This overhead is incurred only during
training; inference is identical to standard RVQ.
\section{Experiments}

\subsection{Architecture}

We build on DAC~\cite{kumar2023highfidelity}. The encoder uses a 1D convolution (kernel 7, 64 channels) followed by four downsampling blocks with stride factors [2, 4, 8, 8] ($512\times$ total). Each block doubles channels ($64 \to 128 \to 256 \to 512 \to 1024$) and contains three dilated residual units (dilations 1, 3, 9), all with Snake activations~\cite{ziyin2020snake}. The decoder mirrors this structure using transposed convolutions. The quantizer has $L=9$ codebooks of 1024 entries in dimension~8, updated via EMA with decay 0.99. At 44.1~kHz, the frame rate is $44100/512 \approx 86$~Hz, giving a full-rate bitrate of $9 \times \log_2(1024) \times 86 \approx 7.7$~kbps.

\textbf{\method{} configuration.} $K=4$ groups with a 3-2-2-2 stage allocation; three codebooks are assigned to bass to reflect its perceptual and energetic dominance. The band loss weight is $\lambda_{\text{band}} = 5$, the Gaussian floor is $\beta = 0.3$, and both $\mu_k$ and $\sigma_k$ are initialized to divide the mel axis evenly ($\sigma_k = M/(3K) \approx 6.7$ mel bins) and learned jointly with the codec. Group dropout probability is $p_{\text{drop}} = 0.5$.

\subsection{Data}

We train on a diverse mixture spanning music, speech, and general audio. Music sources include MUSDB18-HQ~\cite{rafii2017musdb18}, MTG-Jamendo~\cite{bogdanov2019mtg} (folders 00--89 for training, 90--99 for validation), and an in-house collection of over 100k music tracks covering a wide range of genres and production styles. General audio comes from AudioSet~\cite{gemmeke2017audioset}. Speech is drawn from LibriTTS~\cite{zen2019libritts} train-clean-360. Evaluation uses the MUSDB18-HQ test set, FSD50K~\cite{fonseca2021fsd50k} eval, and LibriTTS test-clean. All models are trained on 1-second random crops, batch size 16 across 4 GPUs, for a maximum of 100 epochs (${\sim}$1M steps).

\subsection{Baselines}

We compare against two baselines. \textbf{DAC}~\cite{kumar2023highfidelity} uses the identical architecture and hyperparameters as \method{} but without band supervision, isolating the effect of the proposed training objective. \textbf{BSCodec}~\cite{wang2025bscodec} implements parallel subband decomposition with four separate encoder--decoder pairs targeting bands 0--1, 1--4, 4--10, and 10--22~kHz, matched to \method{}'s band boundaries. To ensure a fair comparison, the BSCodec encoder and decoder are scaled down so that the total parameter count matches DAC and \method{}. BSCodec represents the alternative design philosophy of explicit architectural separation rather than learned specialization, and directly tests whether harmonic context across bands matters.

\subsection{Metrics}

We report metrics across three complementary dimensions.

\textbf{Time-domain fidelity.} Scale-invariant signal-to-distortion ratio (SI-SDR,~dB$\uparrow$) measures waveform-level reconstruction accuracy, normalized to remove gain differences between reference and reconstruction. It is sensitive to temporal misalignment and phase errors, making it a useful indicator of low-level signal quality.

\textbf{Perceptual distance.} Kernel Audio Distance (KAD)~\cite{chung2025kad} computes a maximum mean discrepancy (MMD) between deep feature distributions of reference and reconstructed audio, correlating well with human quality judgments across diverse content. Unlike SI-SDR, KAD is invariant to time-domain phase and captures distributional similarity in a perceptually relevant feature space.

\textbf{Speech intelligibility and quality.} Perceptual Evaluation of Speech Quality (PESQ)~\cite{rix2001perceptual} models the ITU-T P.862 perceptual difference between reference and degraded speech on a 1--4.5 MOS-LQO scale. Short-Time Objective Intelligibility (STOI)~\cite{taal2010short} measures the correlation between short-time temporal envelopes in third-octave bands, providing a direct indicator of speech intelligibility independent of quality.

\textbf{Subjective quality.} We conduct listening tests following the MUltiple Stimuli with Hidden Reference and Anchor (MUSHRA) methodology~\cite{itu2014mushra}. Participants rate each condition on a continuous 0--100 scale (0~=~bad, 100~=~excellent) relative to a hidden reference, allowing fine-grained discrimination between codecs at the same bitrate. MUSHRA scores complement the objective metrics by capturing perceptual dimensions, such as timbral naturalness, bass fullness, harmonic coherence, that SI-SDR and KAD may not fully reflect.

\section{Results}
\label{sec:experiments}

\subsection{Reconstruction Quality}

\begin{table}[t]
  \caption{Reconstruction quality at 7.7~kbps.}
  \label{tab:main}
  \centering
  \small
  \begin{tabular}{@{}llcc@{}}
    \toprule
    \textbf{Domain} & \textbf{Method} & \textbf{SI-SDR}$\uparrow$ & \textbf{KAD}$\downarrow$ \\
    \midrule
    \multirow{3}{*}{Music} & DAC & 8.97 & 0.35 \\
    & BSCodec & 7.96 & 0.30 \\
    & \method{} & $\mathbf{9.22}$ & $\mathbf{0.29}$ \\
    \midrule
    \multirow{3}{*}{Speech} & DAC & 9.75 & 0.19 \\
    & BSCodec & 9.24 & 0.16 \\
    & \method{} & $\mathbf{10.71}$ & $\mathbf{0.14}$ \\
    \midrule
    \multirow{3}{*}{General} & DAC & 7.06 & 0.33 \\
    & BSCodec & 6.52 & 0.37 \\
    & \method{} & $\mathbf{7.39}$ & $\mathbf{0.28}$ \\
    \bottomrule
  \end{tabular}
\end{table}

Table~\ref{tab:main} shows full-bitrate reconstruction. \method{} achieves
the best SI-SDR across all domains. The SI-SDR
gain over DAC is +0.25~dB on music, +0.96~dB on speech, and +0.33~dB on
general audio---modest in absolute terms, but note that DAC and \method{}
share the exact same architecture and parameter count. The only difference
is the training loss. The KAD gains are consistent across all domains:
\method{} achieves the lowest KAD on music ($0.29$ vs.\ 0.30 for BSCodec),
speech ($0.14$ vs.\ 0.16 for BSCodec and 0.19 for DAC), and general audio
($0.28$ vs.\ 0.37 for BSCodec and 0.33 for DAC). The SI-SDR gains over
BSCodec (+1.26~dB music, +1.47~dB speech, +0.87~dB general) are notable
given that BSCodec dedicates a separate encoder--decoder to each band,
yet this architectural specialization is outperformed by cumulative latents,
suggesting cross-band context matters more than band-specific capacity.

\subsection{Spectral Specialization Analysis}

\begin{table}[t]
  \caption{Spectral centroid of each stage group, normalized to $[0,1]$ in mel space.
  \method{} reports learned band centers $\mu_k$ after training;
  DAC reports empirical mel centroids from tier-diff contributions on MUSDB18-HQ test set ($n=50$).}
  \label{tab:spectral}
  \centering
  \small
  \begin{tabular}{lcc}
    \toprule
    \textbf{Group} & \textbf{\method{}} & \textbf{DAC} \\
    \midrule
    0 (Bass)     & 0.08 & 0.06 \\
    1 (Low-mid)  & 0.08 & 0.17 \\
    2 (High-mid) & 0.65 & 0.18 \\
    3 (Treble)   & 0.81 & 0.18 \\
    \bottomrule
  \end{tabular}
\end{table}

Table~\ref{tab:spectral} shows the mel-space centroid of each group's contribution.
\method{}'s learned centers for groups~0 and~1 both converge to 0.08 ($\sim$228~Hz),
indicating the model consolidated its five bass--low-mid codebooks in the fundamental
frequency range of most instruments rather than splitting at the initialized boundary.
Groups~2 and~3 separate clearly at 0.65 and 0.81 ($\sim$5.6~kHz and $\sim$11~kHz),
covering mid and high frequencies as intended.

DAC presents a different picture: tiers~1--3 all land between 0.17 and 0.18
($\sim$575--620~Hz), with no tier reaching above 0.18. Additional codebooks do not
extend spectral coverage upward --- they refine the same low-frequency residual.
This confirms that DAC's spectral entanglement is not merely that each stage captures
a mix of frequencies, but that high-frequency content is essentially absent from
the tier-diff contributions entirely.

Mel-spectrograms of each group's increment
$\hat{x}_k$ (Eq.~\eqref{eq:xk}) on held-out music corroborate this:
group~0 concentrates energy below ${\sim}$1~kHz, groups~1--2 progressively fill
the mid frequencies, and group~3 activates above 10~kHz, whereas DAC's
tier-diffs show broadband energy at every stage.

\subsection{Harmonic Coherence}

\begin{table*}[t]
  \caption{Harmonic reconstruction on 500 synthetic tones ($f_0 = 50$--$200$~Hz, 20 overtones),
  evaluated under globally aligned phases (left) and independent per-harmonic random phases (right).
  BSCodec's elevated variance in the aligned condition and sharp degradation under random phases
  both reflect cross-band incoherence: its per-band decoder has no visibility into adjacent bands
  during reconstruction. \textbf{Bold}: best; \underline{underline}: second best.}
  \label{tab:harmonics}
  \centering
  \small
  \setlength{\tabcolsep}{5pt}
  \begin{tabular}{@{}lcccc@{}}
    \toprule
    & \multicolumn{2}{c}{\textbf{Aligned phases}} & \multicolumn{2}{c}{\textbf{Random phases}} \\
    \cmidrule(lr){2-3} \cmidrule(lr){4-5}
    \textbf{Method} & \textbf{Phase Coh.}$\uparrow$ & \textbf{Amp. RMSE}$\downarrow$
                    & \textbf{Phase Coh.}$\uparrow$ & \textbf{Amp. RMSE}$\downarrow$ \\
    \midrule
    DAC       & $\underline{0.967 \pm 0.012}$ & $\underline{1.93 \pm 0.52}$~dB & $\mathbf{0.987 \pm 0.025}$ & $\mathbf{3.42 \pm 1.48}$~dB \\
    BSCodec   & $0.914 \pm 0.119$ & $2.75 \pm 1.33$~dB & $0.831 \pm 0.173$ & $8.93 \pm 3.60$~dB \\
    \method{} & $\mathbf{0.988 \pm 0.052}$ & $\mathbf{1.83 \pm 0.61}$~dB
              & $\underline{0.986 \pm 0.049}$ & $\underline{4.03 \pm 2.47}$~dB \\
    \bottomrule
  \end{tabular}
\end{table*}

Two stimulus conditions are used (Table~\ref{tab:harmonics}). In the
aligned-phase condition, 500 tones with $f_0 = 50$--$200$~Hz and 20
overtones cross both the 1~kHz and 4~kHz band boundaries. \method{} achieves
the highest phase coherence ($0.988$) and lowest amplitude RMSE
($1.83$~dB), with DAC close behind on phase coherence
($\underline{0.967}$). BSCodec's high variance ($0.914 \pm 0.119$) reveals
selective failure depending on whether a band boundary falls within the
harmonic series.

The random-phase condition sharpens this diagnosis by assigning each
harmonic an independent random phase, making cross-band inference impossible
for a band-isolated decoder. DAC leads on both metrics
($0.987$; Amp.\ RMSE $3.42$~dB), confirming that its shared
latent naturally preserves arbitrary phase relationships. BSCodec collapses
($0.831 \pm 0.173$; Amp.\ RMSE $8.93$~dB), directly confirming the cross-band
incoherence hypothesis. \method{} remains robust ($\underline{0.986}$;
Amp.\ RMSE $\underline{4.03}$~dB) via the cumulative latent, with DAC's
advantage in this condition attributable to the stop-gradient cost at group
boundaries, as confirmed by the ablation in Table~\ref{tab:ablation}.

\subsection{Bitrate Scalability}

\begin{table}[t]
  \caption{SI-SDR (dB) across bitrates on MUSDB18-HQ (Music), LibriTTS (Speech), and FSD50K (General) test sets.}
  \label{tab:bitrate}
  \centering
  \small
  \setlength{\tabcolsep}{10pt}
  \begin{tabular}{@{}llcccc@{}}
    \toprule
    & & \multicolumn{4}{c}{\textbf{Bitrate (kbps)}} \\
    \cmidrule(l){3-6}
    \textbf{Domain} & \textbf{Method} & \textbf{2.6} & \textbf{4.3} & \textbf{6.0} & \textbf{7.7} \\
    \midrule
    \multirow{3}{*}{Music}
      & DAC       & 3.70 & 6.05 & 7.60 & 8.97 \\
      & BSCodec   & 3.58 & 6.31 & 7.03 & 7.96 \\
      & \method{} & \textbf{4.13} & \textbf{6.80} & \textbf{8.09} & \textbf{9.22} \\
    \midrule
    \multirow{3}{*}{Speech}
      & DAC       & 3.74 & 6.51 & 8.29 & 9.75 \\
      & BSCodec   & 4.20 & 7.14 & 8.83 & 9.24 \\
      & \method{} & \textbf{4.41} & \textbf{7.72} & \textbf{9.40} & \textbf{10.71} \\
    \midrule
    \multirow{3}{*}{General}
      & DAC       & -0.16 & 3.18 & 5.26 & 7.06 \\
      & BSCodec   & 0.46  & 3.95 & 5.49 & 6.52 \\
      & \method{} & \textbf{0.59} & \textbf{4.14} & \textbf{5.93} & \textbf{7.39} \\
    \bottomrule
  \end{tabular}
\end{table}

Table~\ref{tab:bitrate} tracks quality as stage groups are added. The advantage
of \method{} over DAC is most pronounced at the lowest bitrate, averaging
+0.6~dB across domains at 2.6~kbps and narrowing to +0.5~dB
at 7.7~kbps. At 2.6~kbps only 3 stages are active, and \method{} reliably
puts bass in those stages. DAC's early stages capture an unstructured mix
and sometimes omit bass entirely. The gap
shrinks at higher bitrates, which makes sense: with all 9 stages available,
even an unstructured allocation eventually covers the full spectrum.

BSCodec's low-bitrate weakness is different. It must spread limited capacity
across four independent encoders, each starved for bits. \method{} concentrates
capacity in the bass group while the Gaussian floor $\beta$ provides soft coverage elsewhere.

\subsection{Perceptual Evaluation}

\begin{table}[t]
  \caption{MUSHRA scores (0--100) at two bitrates. Median [IQR], $n=12$, 12 items (4 music, 4 speech, 4 general audio). DAC and \method{} evaluated from a single trained model; BSCodec is excluded as a stable checkpoint was not available at the time of the listening study.}
  \label{tab:mushra}
  \centering
  \small
  \begin{tabular}{@{}lcc@{}}
    \toprule
    \textbf{Condition} & \textbf{4.3~kbps} & \textbf{7.7~kbps} \\
    \midrule
    DAC       & 59 [43--73] & 84 [68--98] \\
    \method{} & \textbf{68 [50--80]} & \textbf{90 [75--100]} \\
    \midrule
    Reference & 95 [82--97] & 98 [97--99] \\
    \bottomrule
  \end{tabular}
\end{table}

\begin{figure}[t]
  \centering
  \includegraphics[width=\columnwidth]{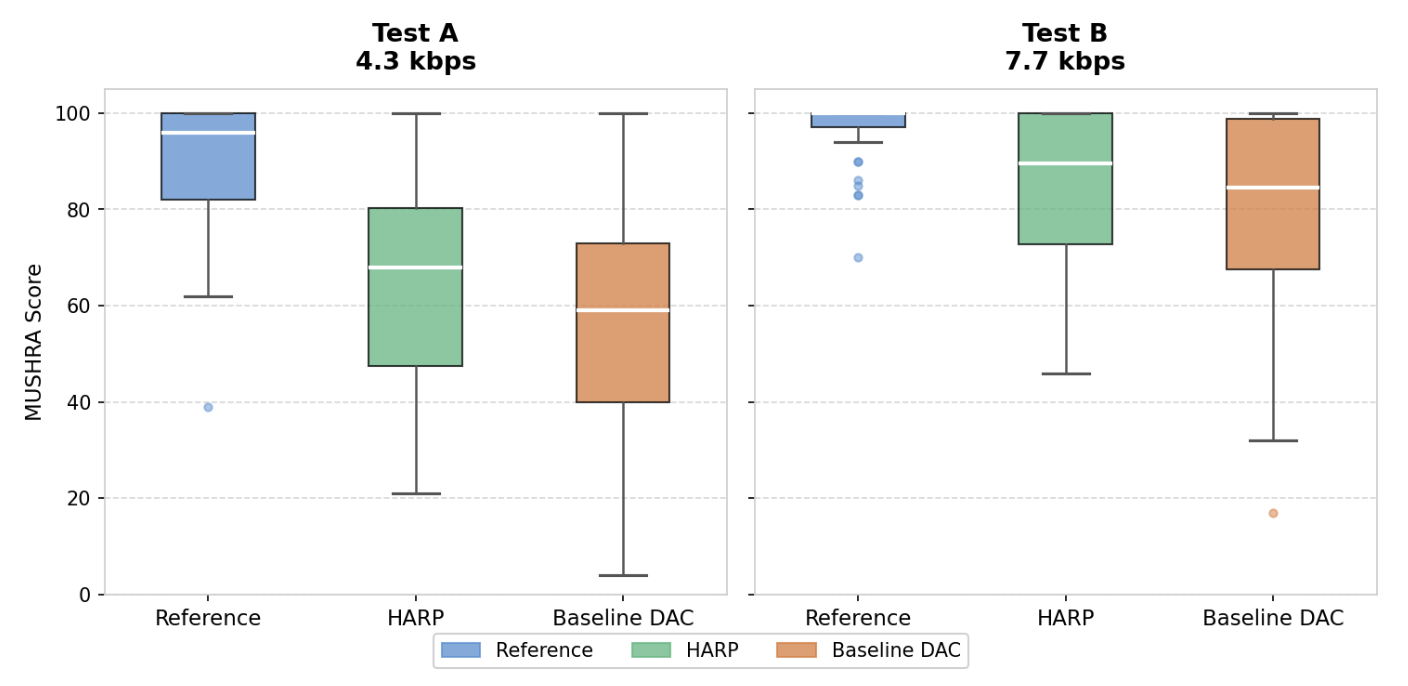}
  \caption{MUSHRA score distributions at 4.3~kbps (Test~A) and 7.7~kbps (Test~B). Boxes show median and IQR; whiskers extend to 1.5$\times$IQR; outliers shown as points.}
  \label{fig:mushra}
\end{figure}

We ran MUSHRA listening tests~\cite{itu2014mushra} with 12 participants rating 12 ten-second items. DAC and \method{}
were evaluated from a \emph{single trained model} at 4.3~kbps (stages 1--5) and
7.7~kbps (all stages), reflecting the adaptive streaming scenario.

\method{} scores significantly above DAC at both bitrates. The advantage is larger at low bitrate
(4.3~kbps: $+9$; 7.7~kbps: $+6$), mirroring the SI-SDR trend in
Table~\ref{tab:bitrate}. DAC's IQR at 4.3~kbps (43--73) is notably wider than
\method{}'s (50--80), indicating that unstructured stage allocation degrades
inconsistently across content types. At 7.7~kbps the gap narrows but remains
perceptually salient (90 vs.\ 84).

Table~\ref{tab:speech} provides speech-specific perceptual scores. \method{}
improves PESQ over DAC by +0.09, consistent with cumulative decoding
preserving the fundamental-to-formant relationship that a flat RVQ does not enforce.
STOI is marginally lower (0.954 vs.\ 0.956, not significant): intelligibility is
envelope-driven and insensitive to the harmonic gains \method{} provides.

\begin{table}[t]
  \caption{Speech perceptual quality on LibriTTS test-clean at 7.7~kbps. KAD reported in Table~\ref{tab:main}.}
  \label{tab:speech}
  \centering
  \small
  \begin{tabular}{@{}lcc@{}}
    \toprule
    \textbf{Method} & \textbf{PESQ}$\uparrow$ & \textbf{STOI}$\uparrow$ \\
    \midrule
    DAC       & $3.167$ & $\mathbf{0.956}$ \\
    BSCodec   & $3.08$  & $0.933$ \\
    \method{} & $\mathbf{3.254}$ & $0.954$ \\
    \bottomrule
  \end{tabular}
\end{table}

\subsection{Ablation Studies}

\begin{table}[t]
  \caption{Ablation on LibriTTS at 7.7~kbps (SI-SDR, dB).}
  \label{tab:ablation}
  \centering
  \small
  \begin{tabular}{lcc}
    \toprule
    \textbf{Configuration} & \textbf{SI-SDR} & $\Delta$ \\
    \midrule
    Full \method{}                    & $\mathbf{10.71}$ & --- \\
    \midrule
    No cumulative decoding            & 9.3  & $-1.4$ \\
    DAC (no band supervision)         & 9.75 & $-1.0$ \\
    Rectangular bands ($\beta = 0$)     & 10.3 & $-0.4$ \\
    No stop-gradient                  & 10.4 & $-0.3$ \\
    \bottomrule
  \end{tabular}
\end{table}

Table~\ref{tab:ablation} isolates the primary design choices, ordered by impact.
Removing cumulative decoding
($-1.4$~dB) eliminates cross-band context entirely, reducing each group to
independent supervision. DAC, which removes all band supervision while retaining
the same architecture, falls $-1.0$~dB below \method{}. Replacing soft Gaussian boundaries
with rectangular ones ($\beta{=}0$, zero weight outside the target band; $-0.4$~dB) has a small but consistent cost, suggesting
the primary benefit of the band loss is spectral guidance rather than the soft
overlap per se. Removing the stop-gradient at group boundaries ($-0.3$~dB) has the
smallest impact, though its effect is consistent with the modest phase coherence
gap over DAC observed in Table~\ref{tab:harmonics}.

\section{Conclusion}

\method{} demonstrates that spectral structure can be imposed on RVQ through
the training loss alone, with no architectural changes. Partitioning stages
into frequency-ordered groups and supervising each group's contribution with
soft, learnable band weights produces codebooks that specialize by frequency,
while cumulative decoding ensures that higher bands are always reconstructed
in the context of their harmonic foundation. The result is consistent
improvements over both standard RVQ and parallel band decomposition across
speech, music, and general audio, with the largest gains at low bitrates
where a structured frequency hierarchy matters most. Inference cost is
unchanged. We hope \method{} offers a practical path toward closing the gap
between the psychoacoustic principles of classical audio coding and the
flexibility of learned compression.

\section{Generative AI Use Disclosure}

Large language models were used during the preparation
of this manuscript for editing and polishing purposes: revising prose for
clarity and academic register, checking mathematical notation for
consistency, and proofreading \LaTeX{} formatting. All technical content,
experimental design, results, and conclusions are the work of the authors.
The authors take full responsibility for the accuracy and integrity of the
manuscript.

\bibliographystyle{IEEEtran}
\bibliography{references}

\end{document}